\documentclass[aip,jcp,reprint,amsmath,amssymb,floatfix,citeautoscript,nofootinbib]{revtex4-2}
\usepackage{cancel}
\usepackage{amsmath}
\usepackage{physics}
\usepackage{graphicx}
\usepackage{amssymb}
\usepackage{amsthm}
\usepackage{bm}

\usepackage{ragged2e}
\usepackage{txfonts}
\allowdisplaybreaks

\usepackage{color}
\usepackage[usenames,dvipsnames]{xcolor}
\definecolor{myblue}{rgb}{0,0,1}
\usepackage[breaklinks=true,colorlinks=true,linkcolor=myblue,urlcolor=myblue,citecolor=myblue]{hyperref}

\usepackage{tabularx}

\begin{document}

\title{
Core binding energies of solids with periodic EOM-CCSD
}

\author{Ethan A. Vo}
\affiliation{Department of Chemistry, Columbia University, New York, NY 10027 USA}
\author{Timothy C. Berkelbach}
\email{t.berkelbach@columbia.edu}
\affiliation{Department of Chemistry, Columbia University, New York, NY 10027 USA}
\affiliation{Initiative for Computational Catalysis, Flatiron Institute, New York, NY 10010 USA}

\begin{abstract}
We report the core binding energies of K-edge and L-edge transitions in simple
semiconducting and insulating solids using periodic equation-of-motion
coupled-cluster theory with single and double excitations (EOM-CCSD).  In our
all-electron calculations, we use triple zeta basis sets with core correlation,
and we sample the Brillouin zone using up to $4\times 4\times 4$ $k$-points.
Our final numbers, which are obtained through composite corrections and
extrapolation to the thermodynamic limit, exhibit errors of about 2~eV when
compared to experimental values. This level of accuracy from CCSD is about the
same as it is for molecules. A low-scaling approximation to EOM-CCSD performs
marginally worse at lower cost, with errors of about 3~eV.
\end{abstract}

\maketitle

\section{Introduction}

X-ray photoelectron spectroscopy measures the binding energies of core
electrons in molecules and solids~\cite{Stevie2020}. The core binding energy
(CBE) of an atom is dependendent on its environment and thus provides a
sensitive probe of local structure. Compared to valence excitations, core
excitations are challenging to predict computationally due to the significant
orbital relaxation in response to the creation of a core hole.  This large
orbital relaxation violates the assumptions of Koopmans' theorem, such that the
mean-field core orbital energy of the neutral system is a very bad
approximation.  The primary methods to predict CBEs in molecules are variants
of the $\Delta$-SCF method that enforce non-Aufbau
occupations~\cite{Chong1995,Besley2009}, the GW approximation to the
self-energy~\cite{Golze2018,Li2022}, the algebraic diagrammatic construction
(ADC)~\cite{Plekan2008,Dreuw2014,Banerjee2019}, and equation-of-motion
coupled-cluster (EOM-CC) theory~\cite{Coriani2015,Vidal2019,Liu2019}.  In
solids, only the first two have been used
much~\cite{Aoki2018,Zhu2021,Kahk2021,Kahk2023}, and a very recent preprint
reports the first such application of ADC~\cite{Ahmed2025}.

Recently, our group and others have reported valence excitations energies of
solids using periodic EOM-CC theory with single and double excitations
(EOM-CCSD)~\cite{McClain2017,Wang2020,Vo2024,Moerman2025,Moerman2025a}. These
works have suggested that EOM-CCSD yields fundamental band gaps and optical
gaps of simple semiconductors, insulators, and color centers with errors of
less than 0.5~eV compared to experimental values, although errors as large as
1~eV are observed~\cite{Moerman2025a} and finite-size errors needs to be
carefully considered~\cite{Moerman2025}.  This performance is on par with
state-of-the-art many-body approaches, such as the GW approximation for band
gaps and the associated Bethe-Salpeter equation for neutral excitations, and
one can imagine inclusion of selected triple excitations towards systematic
improvement.  In this work, we continue this general research agenda by using
EOM-CCSD to calculate the CBEs of several well-characterized solids. As a point
of reference, previous benchmark studies of EOM-CCSD have observed average
errors of about 2~eV in the K-edge CBEs of molecules~\cite{Liu2019}. 

\begin{figure}[b]
	\includegraphics[scale=0.85]{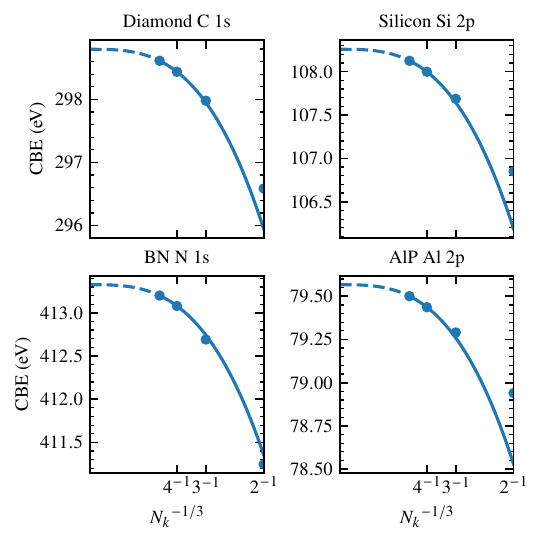}
	\caption{
Convergence of the Hartree-Fock core binding energy (CBE) to the thermodynamic
limit four four of the six transitions studied in this work.  Extrapolation,
shown as a dashed line, is performed assuming finite-size errors that decay as
$N_k^{-1}$.
}
	\label{fig:hf}
\end{figure}

\begin{figure*}
	\includegraphics[scale=0.85]{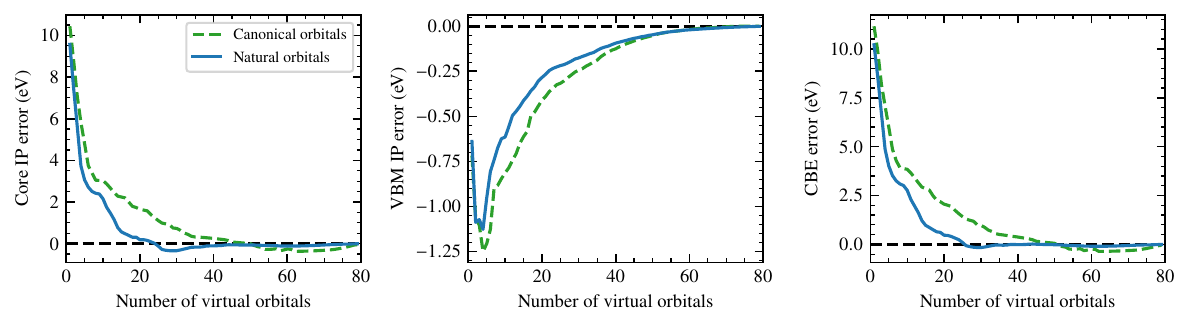}
	\caption{
Convergence of the core IP, valence band maximum (VMB) IP, and core binding
energy (CBE) with the number of correlated virtual orbitals, comparing
canonical and MP2 natural orbitals.  Results are shown for the C 1s CBE in
diamond using $N_k=2^3$ $k$-points.
}
	\label{fig:nos}
\end{figure*}

\section{Methods and Results}

Using PySCF~\cite{Sun2017pyscf,Sun2020}, we performed periodic EOM-CCSD
calculations with $k$-point sampling to calculate ionization potentials (IPs),
as described in Refs.~\onlinecite{McClain2017,Vo2024}. Because PySCF uses
atom-centered Gaussian basis functions, core IPs are straightforward to
calculate without pseudopotentials. We use Gaussian density
fitting~\cite{Sun2017} and the all-electron cc-pCVTZ basis set (augmentation
with diffuse functions is less important in solids than in molecules due to the
significant borrowing of basis functions and the absence of a surrounding
vacuum).

The computational cost of periodic EOM-CCSD calculation is dominated by the
ground-state calculation, which scales as $N_k^4 n_\mathrm{occ}^2
n_\mathrm{vir}^4$ where $N_k$ is the number of $k$-points sampled from the
Brillouin zone, and $n_\mathrm{occ}$ and $n_\mathrm{vir}$ are the number of
occupied and virtual orbitals per $k$-point.  The IP step has a reduced
iterative scaling of $N_k^3 n_\mathrm{occ}^3 n_\mathrm{vir}^2$.  Because core
IPs correspond to interior eigenvalues, we employ the core-valence separation
(CVS) approximation~\cite{Vidal2019,Liu2019} in all core IP calculations.  The
CVS approximation marginally lowers the cost of a matrix-vector product, but
primarily simplifies the use of iterative eigensolvers, such as the Davidson
algorithm used in this work.  Our testing indicates that the CVS approximation
introduces errors of 0.1~eV or less in the solid-state transitions studied
here.

We study five simple semiconductors and insulators (Si, SiC, AlP, C, and cubic
BN), and we calculate the 1s core binding energy (CBE) (K-edge) of the first
row main group elements (C, B, N) and the 2p CBE (L-edge) of the second row
elements (Si, Al).  To determine the CBE, we calculate two IPs, and we report
the CBE referenced to the valence band maximum (VBM),
\begin{equation}
\mathrm{CBE} = \mathrm{IP}_\mathrm{core} - \mathrm{IP}_\mathrm{VBM}.
\end{equation}
All calculations are performed with $k$-point sampling, but IPs are evaluated
at the $\Gamma$ point, which is where the VBM occurs in all materials we study.

In Fig.~\ref{fig:hf}, we present the Hartree-Fock (HF) CBE as a function of the
number of $k$-points sampled in the Brillouin zone (up to $N_k=5^3$), obtained
from the difference in HF orbital energies.  Results are shown for four out of
six of our studied transitions.  Due to an integrable divergence in the
nonlocal exchange~\cite{Sundararaman2013}, the HF orbital energies exhibit an
asymptotic finite-size error that decays as $N_k^{-1/3}$ when using the Ewald
potential~\cite{Fraser1996}, i.e., neglecting the $G=0$ component of the
Coulomb interaction. A Madelung constant
correction~\cite{Paier2005,Broqvist2009} lowers the finite-size error to one
that decays as $N_k^{-1}$. In either case, because the CBE is an energy
difference between occupied orbital energies, its finite-size error decays as
$N_k^{-1}$, which we use to extrapolate to the thermodynamic limit (TDL).
Compared to experimental values (see below), HF overestimates the CBE by about
10--15~eV, which is expected based on its neglect of orbital relaxation.

Because of the high cost of EOM-CCSD with large basis sets and dense $k$-point
meshes, we compress the virtual orbital space using a truncated set of MP2
natural orbitals (NOs)~\cite{Landau2010,Lange2020}. In Fig.~\ref{fig:nos}, we
show an example of the convergence of $\mathrm{IP}_\mathrm{core}$,
$\mathrm{IP}_\mathrm{VBM}$, and the difference $\mathrm{CBE} =
\mathrm{IP}_\mathrm{core} - \mathrm{IP}_\mathrm{VBM}$, as a function of the
number of virtual orbitals. The example is shown for diamond with a $N_k = 2^3$
$k$-point mesh. As can been, the results converge faster when performed in a
truncated NO basis than in the canonical orbital basis.  The basis set error
(compared to the full TZ basis) is dominated by the core IP, and convergence to
0.2~eV is achieved using only 20--30 virtual orbitals, out of a total of 80.

\begin{figure}[b]
	\includegraphics[scale=0.9]{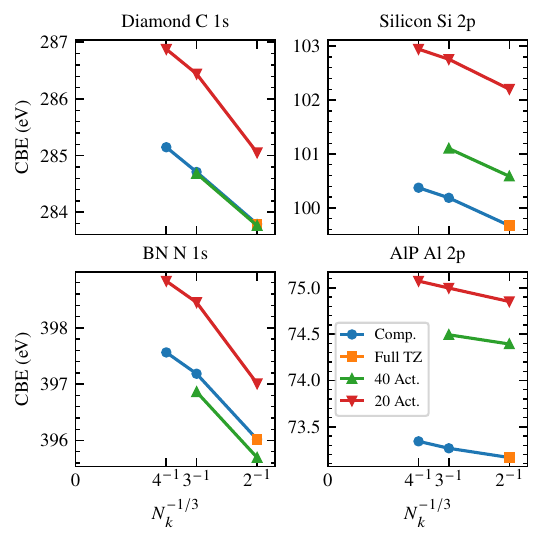}
    \caption{The core binding energy (CBE) as a function of $N_k$ with several
virtual orbital active spaces.  The final composite corrected predictions of
the basis set limit are shown with blue circles.
}
	\label{fig:composite}
\end{figure}

We find that, for each material studied, the convergence behavior with the
number of virtual orbitals included is similar when using denser $k$-point
meshes, suggesting the composite correction, 
\begin{equation} 
E(N_{k,2},\mathrm{L}) 
    \approx E(N_{k,2},\mathrm{S}) + \left[E(N_{k,1},\mathrm{L})-E(N_{k,1},\mathrm{S})\right], 
\end{equation} 
where L and S are large and small virtual orbital spaces.  Specficially, we
combine three sets of calculations: one with the full TZ basis and $N_k= 2^3$,
one with 40 active orbitals (occupied and NO virtuals) up to $N_k = 3^3$, and
one with 20 active orbitals up to $N_k = 4^3$.  Results of this approach are
shown in Fig.~\ref{fig:composite} for the same four transitions.  We see that
an active space containing 40 orbitals has a basis set error of 0.1~eV (C 1s),
0.2~eV (N 1s in BN), 0.8~eV (Si 2p), and 1.2~eV (Al 2p).

\begin{figure}[b]
	\includegraphics[scale=0.9]{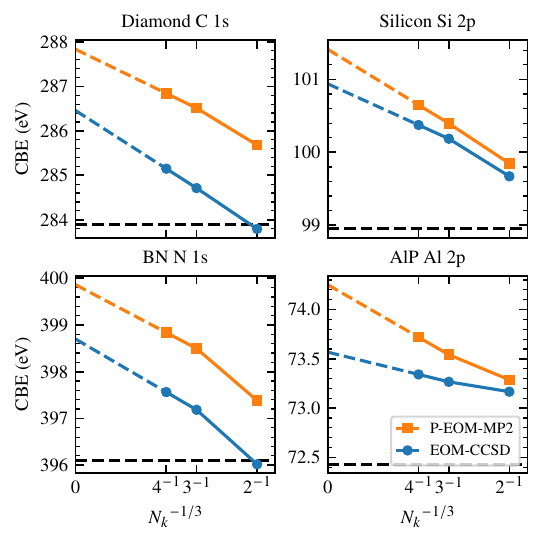}
    \caption{The core binding energy (CBE) as function of $N_k$. The composite
corrected curve EOM-CCSD (blue) is our best estimate of the basis set limit,
which is then extrapolated to the thermodynamic limit. The orange line is a
similar composite corrected curve for P-EOM-MP2. The experimental core binding
energy is indicated by the dashed black line.
}
	\label{fig:cbe_extrap}
\end{figure}

\begin{table*}[t]
	\centering
	\begin{tabular*}{0.98\textwidth}{@{\extracolsep{\fill}} lccccccccc}
\hline\hline
Material & Orbital & Experiment & HF & G$_0$W$_0$@PBE & G$_0$W$_0$@PBE45 & ADC(2) & ADC(2)-X & P-EOM-MP2 & EOM-CCSD \\
		\hline
  Si  & Si 2p & 98.95  & 108.26 &  95.01 &  99.60  & 100.10 &  99.01 &   101.41 & 100.94 \\
 SiC  & C 1s  & 281.45 & 295.99 & 272.71 &  281.67 & 283.56 & 281.33 &   284.67 & 283.52 \\
 AlP  & Al 2p & 72.43  &  79.57 &  68.70 &  72.60  &  73.17 &  72.34 &   74.25  &  73.57 \\
  C   & C 1s  & 283.9  & 298.79 & 276.49 &  284.59 & 286.33 & 283.91 &   287.84 & 286.46 \\
  BN  & B 1s  & 188.4  & 198.31 &        &         & 191.13 & 188.87 &   191.34 & 190.24 \\
      & N 1s  & 396.1  & 413.33 &        &         & 398.00 & 396.36 &   399.86 & 398.70 \\
\hline
MSE (eV)& &            &  12.17 & $-5.95$&    0.43 &   1.84 &   0.10 &     3.02 &   2.03 \\
MAE (eV)& &            &  12.17 &   5.95 &    0.43 &   1.84 &   0.17 &     3.02 &   2.03 \\

		\hline\hline
\end{tabular*}
	\caption{
Core binding energies (CBEs) of the six transitions studied in this work,
comparing results from experiments, HF (this work), 
G$_0$W$_0$~\cite{Zhu2021}, ADC(2) and ADC(2)-X~\cite{Ahmed2025},
and P-EOM-MP2 and EOM-CCSD (this work).  Error statistics are summarized as
mean signed error (MSE) and mean absolute error (MAE). Experimental values are
the ones compiled in Ref.~\onlinecite{Zhu2021}, except for those of BN, which
are from Ref.~\onlinecite{Hamrin1970}.
}
	\label{tab:cbe}
\end{table*}

Our best composite-corrected results are shown in Fig.~\ref{fig:cbe_extrap},
along with the results of a reduced scaling approximation to EOM-CCSD, i.e.,
partitioned EOM-MP2 (P-EOM-MP2), which replaces the CCSD ground state with its
MP2 approximation and replaces the doubles-doubles block of the
similarity-transformed Hamiltonian by a diagonal matrix of orbital energy
differences.  This approximation, which is closely related to the strict ADC(2)
and CC2 methods, has a reduced iterative scaling of $N_k^2 n_\mathrm{occ}^3
n_\mathrm{vir}$, and it was recently shown by our group to predict valence band
gaps in surprisingly good agreement with EOM-CCSD~\cite{Lange2021,Vo2024}.
Because we are interested in predictions in the thermodynamic limit of
$N_k\rightarrow \infty$, we perform extrapolations of our data assuming
finite-size errors that decay as $N_k^{-1/3}$.  Recent works have suggested
that CCSD valence excitation energies (IP/EA) with accessible $k$-point meshes
up to $N_k = 4^3$ exhibit finite-size errors that are not in their asymptotic
regime, and subleading corrections are
expected~\cite{Moerman2025,Moerman2025a}.  Based on the difference between our
predictions with $N_k = 4^3$ and extrapolated to the thermodynamic limit, a
conservative error bar due to finite-size errors is 0.5~eV, which we will find
is smaller than the typical error with respect to experimental values.

Our final HF, P-EOM-MP2, and EOM-CCSD predictions are given in
Tab.~\ref{tab:cbe}, where we compare to experimental values and previously
reported values obtained with the GW approximation based on a PBE reference and
on a PBE45 reference (with 45\% nonlocal exchange)~\cite{Zhu2021}, as well as
with the ADC(2) and ADC(2)-X approximations~\cite{Ahmed2025}.  The GW and ADC
calculations are especially fair comparisons, because they were performed using
PySCF in comparable basis sets and $k$-point meshes.

We find that EOM-CCSD predicts CBEs that are in good agreement with experiment,
but always slightly too large, by about 2~eV. The more affordable P-EOM-MP2
method predicts CBEs that are even larger, i.e., too large by about 3~eV.  Both
of these methods significantly outperform the GW approximation with a PBE
reference, which uniformly underestimates CBEs by about 4--9~eV (6~eV on
average).  In contrast, by including a large fraction of nonlocal exchange, the
GW approximation with a PBE45 reference outperforms P-EOM-MP2 and EOM-CCSD,
achieving a mean absolute error (MAE) of only 0.43~eV. Remarkably, the ADC(2)
and ADC(2)-X predictions are very accurate [MAE of 1.84~eV for ADC(2) and
0.17~eV for ADC(2)-X], despite both being more affordable than EOM-CCSD.  On
the slightly larger set of 14 transitions studied in
Ref.~\onlinecite{Ahmed2025}, ADC(2) and ADC(2)-X yield similar performance,
with MAEs of 1.47~eV and 0.44~eV respectively. 

All results presented in this work are from non-relativistic calculations.
However, we have repeated the IP-EOM-CCSD calculations for all transitions with
$\Gamma$-point sampling of the Brillouin zone using the spin-free exact
two-component (X2C) framework~\cite{Liu2021}.  We find that all CBEs are
modifed by 0.2~eV or less.

\section{Discussion}

Interestingly, the 1--2~eV accuracy that we find for CBEs of solids via
EOM-CCSD is the same as that observed for molecules. For example, the authors
of Ref.~\onlinecite{Liu2019} find that the CCSD C 1s IPs in methane and ethane
are 292.3~eV and 292.0~eV, and the experimental values are 290.9~eV and
290.8~eV.  Similarly, the CCSD N 1s IP in ammonia is 407.1~eV, and the
experimental value is 405.6~eV.  Perhaps most interestingly, those authors show
that errors drop to about 0.2~eV when using CCSDT (nonperturbative triple
excitations).  Therefore, assuming the same transferability of performance
between molecules and solids, we can expect that periodic EOM-CCSDT
calculations would provide unprecedented accuracy in CBEs, perhaps rivaling the
precision achievable in experiments. Although it is unlikely that brute-force
periodic EOM-CCSDT calculations can be performed, composite corrections like
those used here may be sufficient.

Finally, it would be interesting and straightforward to extend the present work
to core-level spectral intensities as well as neutral excitations, which have
been implemented and tested for valence
excitations~\cite{McClain2016,Wang2020,Wang2021}. In particular, satellite
features are famously challenging to quantitatively simulate.  Given their
predominant double excitation character, we expect EOM-CCSD will overestimate
their excitation energies, and improving upon this will be a valuable goal.

\section*{Acknowledgements}
This work was supported by the Air Force Office of Scientific Research under
Grant No.~FA9550-21-1-0400 and the Columbia Center for Computational
Electrochemistry. We acknowledge computing resources from Columbia University’s
Shared Research Computing Facility project, which is supported by NIH Research
Facility Improvement Grant 1G20RR030893-01, and associated funds from the New
York State Empire State Development, Division of Science Technology and
Innovation (NYSTAR) Contract C090171, both awarded April 15, 2010.

\section*{Data Availability Statement}
The data that support the findings of this study are available from the
corresponding author upon reasonable request.

\end{document}